\documentclass{iopjournal}
\definecolor{blueUP}{RGB}{20,25,70}
\usepackage{float}
\usepackage{comment}
\usepackage{amsmath}
\usepackage{bm}
\usepackage{braket}
\usepackage{subcaption}
\usepackage{xcolor}
\usepackage{tikz}
\usepackage{circuitikz}
\usetikzlibrary{arrows,shapes,positioning}
\usetikzlibrary{arrows,shapes,positioning}
\usetikzlibrary{decorations.markings}
\tikzstyle arrowstyle=[scale=1]
\tikzstyle directed=[postaction={decorate,decoration={markings,
    mark=at position .65 with {\arrow[arrowstyle]{stealth}}}}]
\tikzstyle reverse directed=[postaction={decorate,decoration={markings,
    mark=at position .65 with {\arrowreversed[arrowstyle]{stealth};}}}]
\usepackage[percent]{overpic}

\newcommand{\dd}{\text{d}}

\newcommand{\ee}{\text{e}}

\newcommand{\p}{\partial}

\newcommand{\be}{\text{\bf e}}
\newcommand{\eps}{\varepsilon}

\newcommand{\bv}{\text{\bf v}}

\newcommand{\bsigma}{\boldsymbol{\sigma}}

\newcommand{\bnabla}{\boldsymbol{\nabla}}

\definecolor{tissueA}{RGB}{73,167,167}
\definecolor{tissueB}{RGB}{203,25,25}

\begin{document}

\articletype{Paper} 

\title{Interface dynamics in tissue invasion}

\author{Nino Despeignes$^{1*}$\orcid{0000-0000-0000-0000}, Lila Sarfati$^1$\orcid{ 0000-0001-6890-4581} , Marc Durand$^{1}$\orcid{0000-0002-6619-1466
} and Frédéric van Wijland$^{1,2}$\orcid{0000-0002-3335-8573}}

\affil{$^1$Laboratoire Mati\`ere et Syst\`emes Complexes, Université Paris Cité  \& CNRS (UMR 7057), 75013 Paris, France}

\affil{$^2$Yukawa Institute for Theoretical Physics, Kyoto University, Kyoto, 606-8502, Japan}

\affil{$^*$Author to whom any correspondence should be addressed.}

\email{nino.despeignes@u-pariscite.fr}


\begin{abstract}
 We rely on a hydrodynamic description of living tissues to describe the interface separating two of them with distinct constitutive properties. Using the difference in their homeostatic pressures as a control parameter, we show that the interface generates an emergent capillary surface tension that depends on the hydrodynamic scale, and that makes it very stable to large wavelengths perturbations. Using the difference of active forces the two tissues experience as a control parameter, we not only find that the front propagation mechanism shifts from the pushed wave to the Burgers wave, but we also find that the emergent surface tension is not sufficient to stabilize the interface at large enough drive and low enough viscosity.\\

\end{abstract}

\section{The interface between two tissues}
Tissue competition governs key biological processes ranging from embryonic development~\cite{maruyama2017cell} and tissue renewal~\cite{gogna2015cell} to wound healing~\cite{park2017tissue} and tumor invasion~\cite{vishwakarma2020publisher}, where the ability of one cell population to displace another can determine tissue integrity or pathological progression. Understanding the dynamics of interfaces separating competing tissues is therefore essential for explaining how multicellular organisms maintain robust spatial organization despite continuous cell proliferation, death, and migration. In a 2010 seminal work~\cite{ranft2010fluidization}, a modeling of the competing tissues in terms of elasticity and hydrodynamics incorporating the mechanical regulation of cell proliferation was introduced. It was subsequently shown in \cite{ranft2014mechanically} that an initial state with an interface between two competing tissues leads to a propagating front at finite velocity driven by the imbalance of homeostatic pressures between the two tissues (the homeostatic pressure in a given tissue quantifies the nonlinear birth-death population dynamics of cells of a given type). Based on particle models introduced in \cite{ranft2010fluidization}, explored in the two-dimensional two-competing-tissues configuration, it was shown in \cite{podewitz_interface_2016} that the continuous medium approach provided a very faithful description of the dynamics. In the same work,  the statistical properties of the interface at large nonequilibrium drive were also investigated and  deviations from the elastic scaling towards a KPZ-like scaling were numerically observed. In \cite{williamson_stability_2018} the authors used a simplified hydrodynamic approach to describe the interface and its stability. It was argued that at fixed nonequilibrium drive the interface was always unstable, unless a stabilizing surface tension was introduced by hand. Such an instability was subsequently observed in a parameter regime where the tissues compressibilities are sufficiently far apart~\cite{buscher_instability_2020}. In a set of recent  experiments~\cite{Bonnet_collective_2018} involving oncogene-driven tissue competition, the front propagation was explicitly characterized. Interestingly, the front propagation equation introduced in \cite{ranft2014mechanically} also led to more mathematical works aiming at better predicting how the front velocity depends on the various parameters~\cite{pulido2025analysis,campos2025biomechanical}. In addition, active tissue-dependent motility forces were introduced \cite{williamson_stability_2018}. These forces, when unequal, act as an alternative active mechanism driving the invasion of one tissue into another. We derive the hydrodynamics for this particular drive and find that the front propagation and selection mechanism notably departs from the analysis of \cite{ranft2014mechanically}.\\

The concept of surface tension is always delicate to manipulate out of equilibrium. It is only in equilibrium that the same number quantifies a free energy cost, a stress anisotropy or a capillary relaxation time. In this work, we use the continuum description as our starting point to derive a linear evolution equation for weak interface deformations. This tells us about the capillary definition of surface tension (for lack of a better word), which is an emergent property of the interface, beyond mechanical considerations. In the absence of an active drive but with a homeostatic pressure imbalance we find that in the regime of pushed fronts (which is the physically relevant one), the interface is always stable even in the absence of a tissue-dependent differential cell adhesion. When the individual cells comprising the tissues are subjected to distinct active forces, the front propagation is governed by a Burgers-like mechanism. We find that, depending on the tissues' viscosity, a sufficient amount of active forcing is able to destabilize the interface. We explore the parameter space around the instability threshold to draw the corresponding phase diagram. \\

We begin, in Section~\ref{sec:hydrohisto}, by setting the stage of the continuum medium description of the two competing tissues. Once this is established, we present in Section~\ref{sec:interface} a systematic method to derive the dynamics of the deformation field of the interface. This leads to a formal expression of the dispersion relation that compares very well with direct simulations of the hydrodynamics in Section~\ref{sec:simus}. Section~\ref{sec:act} explores not only how the propagation mechanism and how the interface dynamics are affected when each species comprising the tissue is subjected to an active drive. The final section gathers some research directions of interest.

\section{Hydrodynamic description of two competing tissues}
\label{sec:hydrohisto}
Following \cite{ranft2014mechanically}, we model the system as a two species continuous mixture with densities 
$n_i$, $i=A,B$ in three dimensions, with one thin dimension representing the thickness of the tissue. The corresponding velocity field for species $i$ is denoted by $\bv_i$. Cell division and apoptosis is modeled by a deterministic logistic source term:
\begin{equation}
    \partial_t n_i + \bnabla\cdot(n_i \bv_{i}) = k_i n_i
    \label{continuity_eq}
\end{equation}
where the coefficient $k_i$ is the net division rate in the tissue of type $i$. Introducing the local fraction $\phi$ of $A$ particles, we define a average cell velocity 
\begin{equation}
    \bv=\phi\bv_A+(1-\phi)\bv_B
\end{equation}
and an average $A$ flux ${\bf J}=\phi(\bv_A-\bv)$. The evolution equation for $\phi$ then reads~\cite{ranft2014mechanically}
\begin{equation}
    \p_t\phi+\bv\cdot\bnabla\phi=-\bnabla\cdot{\bf J}+(k_A-k_B)\phi(1-\phi)\label{eq_maitressse}
\end{equation}
This equation is supplemented by several phenomenological ingredients. To begin with, the net division rate $k_i$ is regulated by how the pressure field $P$ deviates from the homeostatic pressure $P_i^h$ at which cell death and birth balance in tissue $i$ according to a linear response
\begin{equation}
    k_i \simeq \kappa_i \left( P_i^{h} - P \right)
    \label{eq_close_to_homeostasis}
\end{equation}
where $\kappa_i$ is akin to a  compressibility.  Following \cite{ranft2014mechanically} we take identical compressibility in both tissues (we refer to the appendix~\ref{subsec:app-epsilon} for a discussion of the extension of the hydrodynamic description to unequal compressibilities). The second phenomenological input is the assumption of a Fick's law~\cite{ranft2014mechanically} for ${\bf J}$, namely
\begin{equation}
    {\bf J}=-D\bnabla \phi
\end{equation}
Finally, the attention goes to the local stress tensor $\bsigma$ which, in the thin film approximation, and after integrating along the third (thin) dimension, balances out the viscous drag:
\begin{equation}
    \bnabla\cdot\bsigma = \gamma \bv
    \label{eq_overdamped}
\end{equation}
Note that the operator $\bnabla$, the stress tensor and all vector fields are now restricted to the two-dimensional space transverse to the thin direction.

This equation connects the velocity field to the mechanics of the tissue, but a final step is needed to close the system of equations. The stress tensor splits into a diagonal pressure term and a traceless term involving the Ericksen contribution:
\begin{equation}\label{eq_stress}
    \sigma_{\alpha\beta}=-P\delta_{\alpha\beta}+2 \eta\, \tilde{v}_{\alpha\beta}
    - B \left(
    (\partial_\alpha \phi)(\partial_\beta \phi)
    - \frac{1}{2}(\bm \nabla \phi)^2\delta_{\alpha\beta}
    \right)
\end{equation}
where $ \tilde{v}_{\alpha \beta} = \frac{1}{2}\left(\partial_{\alpha}v_{\beta} + \partial_{\beta}v_{\alpha} -  \bm \nabla \cdot \bv  \delta_{\alpha \beta} \right)$ and where $\eta$ is the shear viscosity. The $B$-dependent term, arising from differential cell adhesion, gives rise to an effective line tension for the interface of mechanical origin~\cite{ranft2014mechanically,buscher_instability_2020}. The total volume occupied by both cell populations is constant, which enforces the constraint
\begin{equation}
    \bm \nabla \cdot \bv = \phi k_A + \left( 1-\phi \right)k_B
    \label{eq_incompressibility}
\end{equation}
Equation~ \eqref{eq_close_to_homeostasis} is then used to eliminate the pressure field,
\begin{equation}
    \begin{split}
    P = P_B^h -\frac{\bm \nabla\cdot \bv}{\kappa}  + \phi  P_A^h 
    \end{split}
    \label{eq_Pressure}
\end{equation}
This leaves us with two coupled equations for $\phi$ and $\bv$, namely Eqs.~\eqref{eq_maitressse} and \eqref{eq_incompressibility} in which $P$ (which appears in $k_i$) is substituted with its expression in Eq.~\eqref{eq_Pressure}. We therefore arrive at
\begin{equation}
    \begin{split}
        T_{\alpha\beta}v_{\beta} =& -\frac{\Delta P^h}{\gamma} \partial_{\alpha}\phi -  \frac{B}{\gamma} \left(
        \partial_{\beta}(\partial_{\beta} \partial_{\alpha}\phi)
        - \frac{1}{2}\partial_{\alpha}(\partial_{\gamma} \phi)(\partial_{\gamma} \phi)
        \right)
    \end{split}
    \label{V_eq}
\end{equation}
where the hydrodynamic interaction tensor $T_{\alpha\beta}$ has the expression
\begin{equation}
    T_{\alpha \beta} = \left(1 - \frac{\eta}{\gamma}\Delta \right)\delta_{\alpha \beta} - \frac{1}{\kappa \gamma}\partial_{\alpha}\partial_{\beta}
    \label{eq_tenseur_T}
\end{equation}

One final step is needed to cast the formulation of the hydrodynamics in a convenient form: we introduce dimensionless time, space and velocity defined by
\begin{equation}
    t'=t/t_0,\,\, x'=x/\ell_0,\,y'=y/\ell_0,\,\,\, \bv'(x',y',t')=\bv(x,y,t) t_0/\ell_0
    \label{eq_dimensionless}
\end{equation}
Following~\cite{ranft2014mechanically} we choose $t_0=(\Delta P^h \kappa)^{-1}$ and $\ell_0=\sqrt{D/\Delta P^h \kappa}$. Upon dropping   the primes we arrive at the dimensionless equation
\begin{equation}\label{eq_phi}
    \begin{split}
        \partial_t \phi  + {\bf v}\cdot \bm \nabla \phi = \Delta \phi  + \phi\left(1-\phi\right)
    \end{split}
\end{equation}
The linear equation for $\bv$ takes the form
\begin{equation}\label{eq_v-enfonctiondephi}
    \begin{split}
        \begin{pmatrix}
        1 - \Lambda_\perp^2\partial_x^2 - \Lambda_\parallel^2\partial_y^2
        &
        -\left(\Lambda_\perp^2 -\Lambda_\parallel^2\right)\partial_x\partial_y
        \\
        -\left(\Lambda_\perp^2 -\Lambda_\parallel^2\right)\partial_x\partial_y
        &
        1 - \Lambda_\perp^2\partial_y^2 - \Lambda_\parallel^2\partial_x^2
        \end{pmatrix}
        \bv =&  -2\Lambda_\perp V_0 \bm \nabla \phi \bigg(1+ \beta \Delta \phi\bigg)
    \end{split}
\end{equation}
where we have introduced 
\begin{equation}
    \Lambda_\perp= \sqrt{\frac{\Delta P^h \left(1 + \eta\kappa\right)}{D \gamma}}, \;\;\Lambda_\parallel= \sqrt{\frac{\Delta P^h \eta \kappa}{D \gamma}},\;\;V_0= \frac{\Delta P^h}{2D\gamma \Lambda_{\perp}},\;\;\beta= \frac{B \kappa}{D} 
\end{equation}
There exists an explicit relation linking $\Lambda_{\perp}$, $\Lambda_{\parallel}$ and $V_0$, which defines the domain of physical parameters:
\begin{equation}
    \begin{split}
        \Lambda_{\perp} = V_0 + \sqrt{V_0^2 + \Lambda_{\parallel}^2}
    \end{split}
    \label{eq_physical_params}
\end{equation}
The only physical $\Lambda_{\perp}$ values are the one greater than $2 V_0$. If $V_0 \neq 0$,equation \eqref{eq_physical_params} constrains $\Lambda_{\perp} \neq \Lambda_{\parallel}$ with $\Lambda_{\perp}^2 - \Lambda_{\parallel}^2 = 2\Lambda_{\perp} V_0$. The analysis detailed below holds in the case of pushed fronts, so that we further restrict the domain of $(\Lambda_{\perp},V_0)$ to be in that regime~\cite{ranft2014mechanically}.
The combined Eqs.~\eqref{eq_phi} and \eqref{eq_v-enfonctiondephi} are the starting point of our analysis. They differ from those of \cite{ranft2014mechanically, campos2025biomechanical} in several respects. First, they retain the two dimensional nature of the tissue. As a related consequence, the velocity is a vector field. Second, they feature one extra hydrodynamic length scale $\Lambda_\parallel$.

\section{Deriving an equation for the interface}
\label{sec:interface}
\subsection{Critical discussion of the state of the art}
In the original work~\cite{ranft2014mechanically} the authors focus on the front separating the two tissues described as a flat interface translationally invariant in the direction $y$ perpendicular to the propagation of the front (along $x$). The resulting effectively one-dimensional non linear partial differential equation for $\phi(x,t)$ and $v_x(x,t)$ conceals a very rich phenomenology of pulled and pushed fronts that has recently been the subject of further mathematical investigations~\cite{campos2025biomechanical}. In an effort to explore the properties of the interface, two routes have been followed.\\

As we have mentioned in our introduction, particle-based simulations have been used to probe the relevance of the continuous description and to investigate not only the stability properties of the interface but also its nonlinear regime.\\

On the analytical side, the approach in \cite{williamson_stability_2018} consists in viewing the propagating front as a sharp interface, namely an impervious boundary between two immiscible domains of $A$'s and $B$'s. This simplifying assumption replaces the nonlinear front velocity selection mechanism by a boundary condition connecting the two tissues. This allows for considerable analytic progress: The Edwards-Wilkinson dispersion relation for the relaxation predicts that the flat interface is stable as long as $\Delta P^h$ remains below a given threshold. This threshold, however, is proportional to a surface-tension-like stress term at the interface between the two tissues characterized by a new phenomenological coefficient. However, the approach of \cite{williamson_stability_2018} also has some limitations. For instance, as can be seen both in experiments~\cite{Bonnet_collective_2018} (Figs. 1b and 2) and  simulations~\cite{podewitz_interface_2016, buscher_instability_2020} (Fig. 2 in both), there is some level of mixing at the interface between the two tissues, which the original hydrodynamic description in \cite{ranft2014mechanically} naturally contains.\\

Perhaps more importantly, the surface tension introduced by hand in \cite{williamson_stability_2018} is a proxy for the Ericksen contribution proportional to $B$ standing for differential cell adhesion in Eq.~\eqref{eq_stress}, as argued in \cite{ranft2014mechanically}. However, out of equilibrium the notion of surface tension is a protean concept~\cite{langford2025mechanics}. The one governing mechanics need not be directly related to the one governing capillary relaxation~\cite{bialke2015negative,fausti2021capillary,besse2023interface,zhao2024active,sarfati_bulk_2026}. And indeed, as we  show below, there is already a nonzero capillary surface tension even when the Ericksen contribution proportional to $B$ in Eq.~\eqref{eq_stress}  vanishes. Importantly, this surface tension depends on the parameters driving the front; it is not an independent coefficient. We now present our derivation of the interface dynamics starting directly from the hydrodynamics.

\subsection{Defining the interface}\label{subsec:homeo}
\begin{figure}[H]
\begin{center}
\begin{tikzpicture}
\node  at (0,0) {\includegraphics[width=8cm]{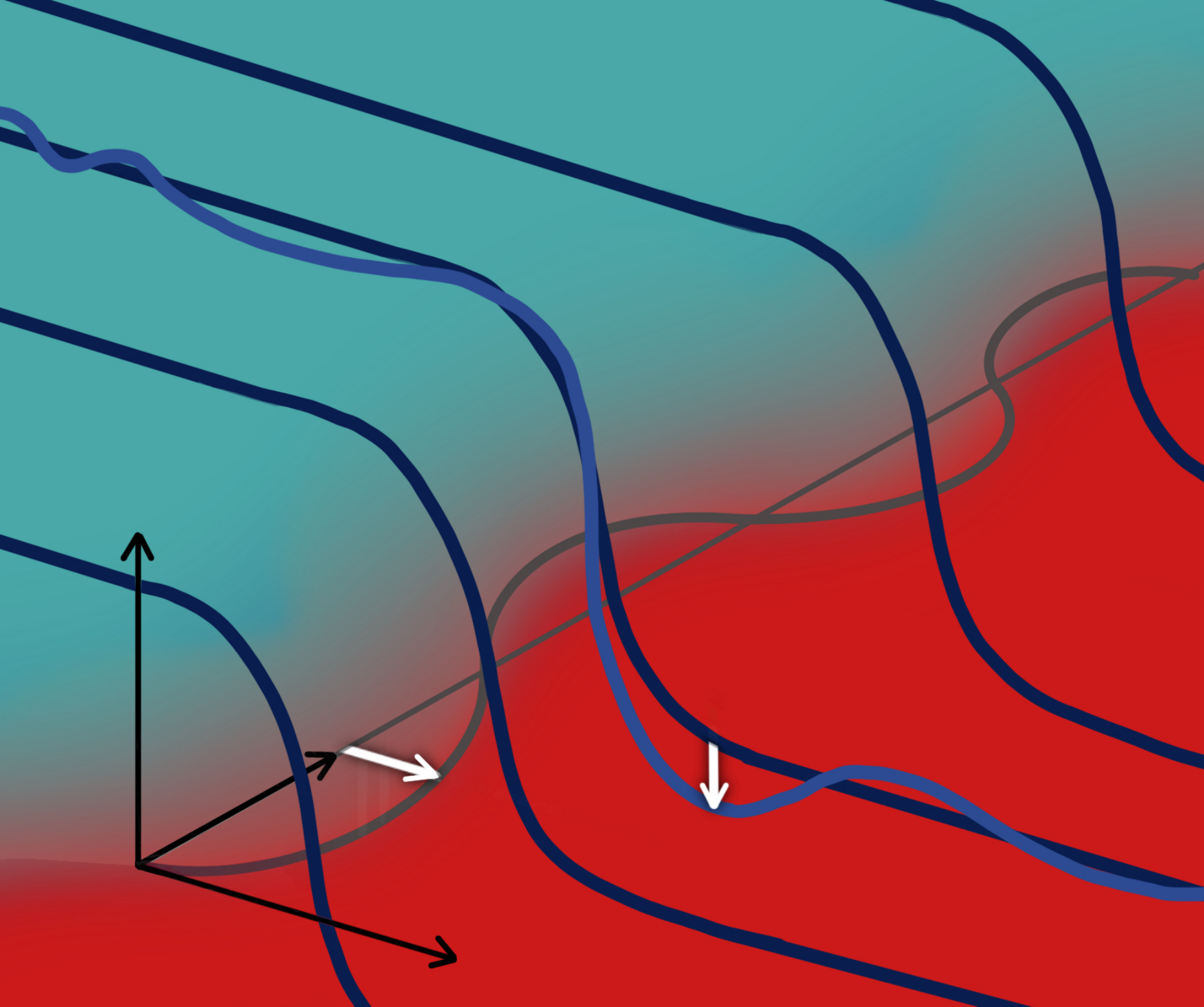}};
\node at (-.8,-3) {$u$};
\node at (-2.5,-1.8) {$y$};
\node at (-3.1,0) {$\textcolor{blueUP}{\Phi(u)}$};
\node at (-0.9,-1.3) [left] {$\textcolor{white}{h(y,t)}$};
\node at (2.2,-1.3) [left] {$\textcolor{white}{\delta\phi(u,y,t)}$};
\node at (-3.5,3) [left] {$\textcolor{white}{A}$};
\node at (4,-3) [left] {$\textcolor{white}{B}$};
\end{tikzpicture}
 \caption{Cartoon of the interface in the comoving reference frame with coordinate $u$. Left is the $\textcolor{tissueA}{A}$ rich phase, and right is the $\textcolor{tissueB}{B}$ rich phase, with $\Delta P^h>0$ leading to an interface effectively propagating to the right ($c>0$). The flat front has a profile $\Phi(u)$ independent of $y$. The algebraic deviation from the flat front is quantified by $h(y,t)$, and deviations of the density with respect to $\Phi$ are described by $\delta\phi$. }
\label{fig:cartoonh}
\end{center}
\end{figure}
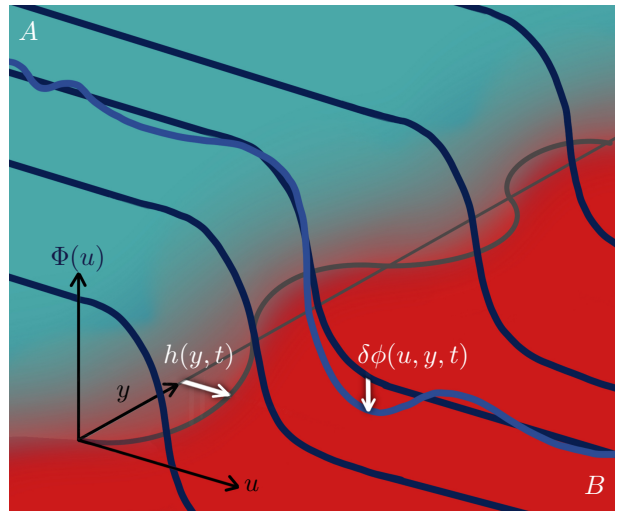

We work in the reference frame of the moving front which has a velocity $c$ along the $x$ direction. The new horizontal coordinate is henceforth  denoted by $u=x-ct$ and the stationary front profile is denoted by $\Phi(u)$. The stationary velocity field is along $u$ only and is denoted by $V(u)$. The interface deformation $h(y,t)$ is defined as the deviation from the flat front, as shown in the cartoon of Fig.~\ref{fig:cartoonh}.
As extensively discussed in \cite{sarfati_bulk_2026}, a definition of the interface deformation $h(y,t)$  used since the eighties~\cite{dashen1974nonperturbative,gervais1975extended,gervais1975perturbation,diehl1980interface,kuramoto1980instability,bausch1981critical,kawasaki1982kinetic2,bausch1991effects}, involves a function $L(u)$ and the deviation $\delta\phi(u,y,t)=\phi(u+h(y,t),y,t)-\Phi(u)$ from the stationary profile such that
\begin{equation}\label{eq_definterface}
    \int\dd u L(u) \delta\phi(u,y,t)=0
\end{equation}
The function $L$ is the zero eigenvalue eigenvector of the adjoint of the linear operator governing the stability of $\Phi$. The difference with the hydrodynamic approach of \cite{williamson_stability_2018} lies in the introduction of the deviation field $\delta\phi$ also depicted in Fig.~\ref{fig:cartoonh}. The latter accounts for deviations within the bulk and it is coupled to the relaxation of $h$. In spite of being a fast field, its relaxation must be considered at given interface profile $h$, which renders its elimination a little cumbersome (hence the introduction of the function $L$). In many ways, and in spite of its nonequilibrium nature, the problem at hand is reminiscent of the interface between two stable phases in an equilibrium model H dynamics in which the order parameter is advected by a velocity field~\cite{kawasaki1982kinetic2,shinozaki1993dispersion,sarfati_bulk_2026}.\\

It is convenient to split the fields $\phi$ and $\bv$ into their shifted stationary profile and a deviation:
\begin{equation}\label{eq_deviations}
   \left\{\begin{array}{l}
        \phi(u,y,t)=\Phi(u-h)+\delta\phi(u-h,y,t)\\v_u(u,y,t)=V(u-h)+\delta v_u(u-h,y,t),\\v_y(u,y,t)=\delta v_y(u-h,y,t)
  \end{array}  \right.
\end{equation}
where $h=h(y,t)$. These relations define the deviations $\delta\phi$ and $\delta\bv$.

For weak deformations of the interface, we can linearize Eqs.~\eqref{eq_phi} and \eqref{eq_v-enfonctiondephi} after plugging the form Eq.~\eqref{eq_deviations}. This leads to three linear yet coupled equations involving the four unknowns $h$, $\delta\phi$ and $\delta\bv$:
\begin{equation}
    \begin{split}
    \left\{\begin{array}{ll}
        &-\partial_t h \partial_u \Phi + \partial_t \delta\phi = \Delta\delta\phi  + (c-V)\partial_u \delta\phi - \partial_u\Phi \delta v_u - \partial_y^2 h \partial_u \Phi \\
        \\
         &\bigg(1 - \Lambda_\perp^2 \partial_u^2 - \Lambda_\parallel^2 \partial_y^2 \bigg) \delta v_u  - \left( \Lambda_\perp^2 - \Lambda_\parallel^2\right)\partial_u\partial_y \delta v_y + \Lambda_\parallel^2 \partial_y^2 h \partial_u V=-2\Lambda_\perp V_0\bigg(\partial_u\delta \phi \\
         &\qquad\qquad\qquad\qquad\qquad\qquad\qquad\qquad\qquad\qquad\qquad\qquad+ \partial_u\Phi\partial_u^2\delta \phi + \partial_u^2\Phi\partial_u \delta \phi + \partial_u\Phi\partial_y^2\delta \phi\bigg)\\
         \\
         &\bigg(1 - \Lambda_\parallel^2 \partial_u^2 - \Lambda_\perp^2 \partial_y^2 \bigg) \delta v_u  - \left( \Lambda_\perp^2 - \Lambda_\parallel^2\right)\left(\partial_u\partial_y \delta v_y -\partial_y h \partial_u^2 V\right)=-2\Lambda_{\perp} V_0 \bigg(\left(\partial_y \delta \phi - \partial_y h\partial_u \Phi\right)\\
         &\qquad\qquad\qquad\qquad\qquad\qquad\qquad\qquad\qquad\qquad\qquad\qquad\qquad\qquad\qquad\qquad\times\big(1 + \beta \p_u^2 \Phi \big) \bigg) \end{array}  \right. 
    \end{split}
\end{equation}
The last two equations can be used to express $\delta\bv$ in terms of $\delta\phi$ and $h$ via  spatial convolutions. Defining $\zeta_{\perp}(q) = \frac{\Lambda_{\perp}}{\sqrt{1+(\Lambda_{\perp}q)^2}}$ and $\zeta_{\parallel}(q) = \frac{\Lambda_{\parallel}}{\sqrt{1+(\Lambda_{\parallel}q)^2}}$, one can obtain explicitly the inverse $\mathcal{G}$ of the linear operator acting on $\delta \bv$.  When written in the Fourier space associated to the $y$-coordinate it is local in $q$ and given by 
\begin{equation}\label{eq_Green}
    \mathcal{G}(q) =
        \begin{pmatrix}
         \frac{\ee^{-\frac{|u|}{\zeta_{\perp}(q)}}}{\zeta_{\perp}(q)} - q^2 \zeta_{\parallel}(q) \ee^{\frac{-|u|}{\zeta_{\parallel}(q)}}
        &
        -iq \text{ sign}(u)\bigg(\ee^{-\frac{|u|}{\zeta_{\perp}(q)}} - \ee^{\frac{-|u|}{\zeta_{\parallel}(q)}}\bigg)
        \\[6pt]
        -iq \text{ sign}(u)\bigg(\ee^{\frac{-|u|}{\zeta_{\perp}(q)}} - \ee^{\frac{-|u|}{\zeta_{\parallel}(q)}} \bigg)
        &
         \frac{\ee^{\frac{-|u|}{\zeta_{\parallel}(q)}}}{\zeta_{\parallel}(q)} - q^2 \zeta_{\perp}(q) \ee^{\frac{-|u|}{\zeta_{\perp}(q)}}
        \end{pmatrix} 
\end{equation}
and thus the first equation bears on $\delta\phi $ and $h$ only. Regardless of this technical difficulty, the general structure of the resulting equation has the form
\begin{equation}
  \Gamma(\delta\phi- h\p_u\Phi) =\Gamma' h\p_u\Phi
  \label{eq_operator}
\end{equation}
where both $\Gamma$ and $\Gamma'$ are linear integro-differential operators whose explicit expressions are collected in the appendix~\ref{subsec:app-gamma}.  The operator $\Gamma$ becomes local in $q$, which we make explicit by   writing $\Gamma(q)$ (this operator acts in the space of $(u,t)$-dependent functions). The operator $\Gamma$ splits into
\begin{equation}
    \Gamma=\p_t + F(q,u) + \Omega
\end{equation}
where $\Omega$ is the operator governing the linear stability of the the profile $\Phi(u)$ with respect to a one-dimensional $u$-dependent perturbation, and $F(q,u)$ is an operator acting in the space of $u$-dependent functions that vanishes when $q=0$ (see also \ref{subsec:app-gamma}. By construction, note that $\Omega$, and thus $\Gamma(q=0)$, have an eigenvalue zero with eigenvector $\p_u\Phi$. We choose $L$ to be the corresponding eigenvector of $\Omega^\dagger$, where the dagger refers to the canonical scalar product in the space of functions of $u$. In order to disentangle $\delta\phi$ from the field of interest $h$, the idea is to operate on both sides of Eq.~\eqref{eq_operator} by $\Gamma^{-1}$, then to multiply by $L$ and finally to integrate over $u$: 
\begin{equation}\begin{split}
 \int_uL\delta\phi- h(q,t)\left(\int_uL\p_u\Phi\right) =&h(q,t)\int_u L\Gamma^{-1}\Gamma' \p_u\Phi\\
 - h(q,t)\left(\int_u L\p_u\Phi\right)=&h(q,t)\int_u\left((\Gamma^{-1})^{\dagger}L\right)\Gamma' \p_u\Phi,
\end{split}\end{equation}
where we have used the definition Eq.~\eqref{eq_definterface} of $h$. In  the long-time regime, the integral on the right hand side is dominated by the eigenvalue $\Lambda_0(q)$ of $F(q,u) + \Omega$ whose real part  is closest to zero; the corresponding eigenvectors are $L_0(q,u)$ and $R_0(q,u)$. Hence, in the $q \to 0$ limit, and using that $R_0(q,u)\to \p_u\Phi$, $L_0(q,u)\to L(u)$ as $q\to 0$, the relevant part of the spectral decomposition of  $(\Gamma^{-1})^{\dagger}$ is
\begin{equation}
     (\Gamma^{-1})^{\dagger}(\omega;q;u,u') = \frac{1}{i\omega + \Lambda_0(q)}\frac{L(u)\partial_{u'}\Phi(u')}{\int_{u''} L(u'') \partial_{u''}\Phi(u'')}
\end{equation}
where we have Fourier transformed in time as well. The eigenvalue $\Lambda_0(q)$ is given to leading order in $q\to 0$ by
\begin{equation}
   \Lambda_0(q) = \frac{\int_u L(u) F(q,u)\partial_u\Phi(u)}{\int_u L \partial_u\Phi}.
\end{equation}
We therefore arrive at a generic linear evolution equation for $h$ of the form
\begin{equation}\label{eq_beauty}
-\left[\int_uL\p_u\Phi\right]\p_t h(q,t)=\left[\int_u L(u)\Gamma'(q,u) \p_u\Phi(u)\right] h(q,t) +\left[\int_u L(u) F(q,u)\p_u\Phi(u)\right] h(q,t)
\end{equation}
Equation~\eqref{eq_beauty} is the central mathematical piece of our physical analysis. It provides a linear evolution equation for $h$ to leading order in the $q\to 0$ limit in which the dispersion relation can be read off from the various $u$ integrals that appear on either side. Equation~\eqref{eq_beauty} can be written in the form
\begin{equation}
    \p_t h(q,t)=-\omega(q) h(q,t),\,\,\omega(q)=\frac{\int_u L(u)\left[\Gamma'(q,u)+F(q,u)\right]\p_u\Phi(u)}{\int_u L\p_u\Phi}.
\end{equation}
In order to proceed, we need the perfect knowledge of $\Phi$ and $L$. While this is often out of analytical reach, it is a relatively simple task to obtain them numerically from the one-dimensional problem. We thus resort to functions obtained by a numerical solution of the corresponding integro-differential equations.\\

This important technical part being sorted out, we are now in a position to reap the fruits of our mathematical effort. We shall begin with investigating the dispersion relation in the $B=0$ case, in order to show the emergence of a nonzero capillary surface tension.  

\section{Stability of the interface}\label{sec:simus}
\subsection{In the absence of the Ericksen stress tensor}
We begin our analysis of the $B=0$ limiting case, the dispersion relation $\omega(q)$ can be investigated in itself. When the hydrodynamic scales $\Lambda_{\perp,\parallel}$ are small, we expect an Edwards-Wilkinson scaling $\omega(q)\propto q^2$. As we now explore, taking into account the hydrodynamics also leads to an Edwards-Wilkinson scaling, albeit with a different proportionality factor. The expression for $\omega(q)$ is given by

\begin{equation}
    \begin{split}
        \omega(q) =& q^2 - \frac{V_0}{2\Lambda_{\perp}\int_u L(u)\p_u\Phi(u)}\int_u L(u) \p_u\Phi(u) \int_{u'}\bigg(\ee^{-|u-u'|/\Lambda_{\perp}}\\
        &- \ee^{-|u-u'|\sqrt{1+(\Lambda_{\perp}q)^2}/\Lambda_{\perp}} \bigg)\text{sign}(u-u')\partial_u'\Phi(u')
    \end{split}
\end{equation}

We now explore the limiting behavior of this dispersion relation in two regimes. In the $ q\xi,\,q\Lambda_{\perp}\,\ll 1$ regime where hydrodynamics plays a role, we have the limiting behavior 
\begin{equation}
    \begin{split}
        \frac{\omega(q)}{q^2}\underset{q\Lambda_{\perp},\,q\Lambda_{\parallel}\to 0}{\longrightarrow}&1-\frac{V_0}{4\int_uL\p_u\Phi}\int_u L(u) \partial_u\Phi(u) \int_{u'}\ee^{-|u-u'|/\Lambda_{\perp}}(u-u')\partial_{u'}\Phi(u')
    \end{split}
\end{equation}
The correction in the right hand side turns out to be positive and greater than 1, as is shown in Fig.~\ref{fig:hydro-cross-EW} where $\omega(q)/q^2$ is plotted for $B=0$ for various values of $\Lambda_{\perp,\parallel}$ and $V_0$ while keeping $q\Lambda_{\perp,\parallel} $ small.\\

In the other limit of interest where hydrodynamics  does not matter, namely when $ q\xi\ll 1\ll q\Lambda_{\perp}$, we arrive to
\begin{equation}
    \frac{\omega(q)}{q^2}\underset{q\Lambda_{\perp}\to\infty}{\longrightarrow}
1
\end{equation}

This shows the existence of a hydrodynamics-driven crossover between two Edwards-Wilkinson regimes as is visible in Fig~.\ref{fig:hydro-cross-EW}.

We therefore conclude that when $B=0$, the interface is described by a stable Edward-Wilkinson equation. As we have shown, the interplay between the logistic growth and the hydrodynamics-driven pushed front effectively endows the interface with a capillary effective surface tension that brings it back to its flat state if  perturbed. This is fully consistent with the stability of the interface found in the agent-based simulations of \cite{podewitz_interface_2016, buscher_instability_2020} which are specifically carried out using symmetric interactions between cells, regardless of their nature (we view this hypothesis of theirs as effectively working with $B=0$).\\

When investigating the stability of the interface, working with $B=0$ is in some sense a worst case scenario for keeping the interface flat. Indeed, as physical intuition dictates, bringing a nonzero $B$ back into the physical description can only further stabilize the interface.

\subsection{Including differential cell adhesion}
The hydrodynamic model allows for considering differential cell adhesion, which reflects in a coefficient $B\neq 0$ in the Ericksen stress tensor. As discussed previously, we anticipate that such an additional ingredient should stiffen, and thus stabilize, the interface. The dispersion relation now reads

\begin{equation}\label{eq:dispersion-gore}
    \begin{split}
         \omega(q) =& q^2 - \frac{\Lambda_{\perp}V_0}{\int_u L\p_u\Phi}\int_u L(u) \p_u\Phi(u) \int_{u'} \bigg\{\frac{1}{\Lambda_{\perp}^2}\bigg(\ee^{-|u-u'|/\Lambda_{\perp}}\\
        &- \ee^{-|u-u'|\sqrt{1+(\Lambda_{\perp}q)^2}/\Lambda_{\perp}} \bigg)\text{sign}(u-u')\partial_u'\Phi(u')\bigg(1+\beta \partial^2_{u'} \Phi(u')\bigg)\\
        &-\beta q^2\bigg(\frac{\ee^{-|u-u'|\sqrt{1+(\Lambda_{\perp}q)^2}/\Lambda_{\perp}}\sqrt{1+(\Lambda_{\perp}q)^2}}{\Lambda_{\perp}} - \frac{\ee^{-|u-u'|\sqrt{1+(\Lambda_{\parallel}q)^2}/\Lambda_{\parallel}}\Lambda_{\parallel}q^2}{\sqrt{1+(\Lambda_{\parallel}q)^2}}\bigg)(\partial_{u'}\Phi(u'))^2\bigg\}
    \end{split}
\end{equation}

\begin{figure}[H]
\centering
\begin{subfigure}{0.49\textwidth}
    \centering
    \includegraphics[width=\linewidth]{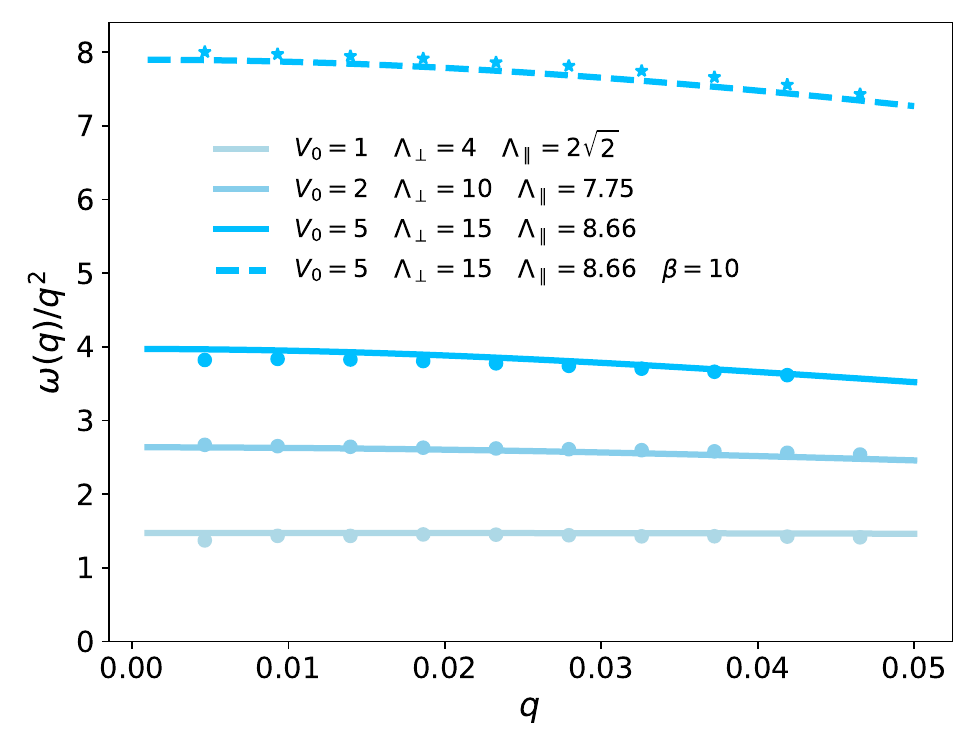}
    \label{fig:graphe1}
\end{subfigure}
\hfill
\begin{subfigure}{0.49\textwidth}
    \centering
    \includegraphics[width=\linewidth]{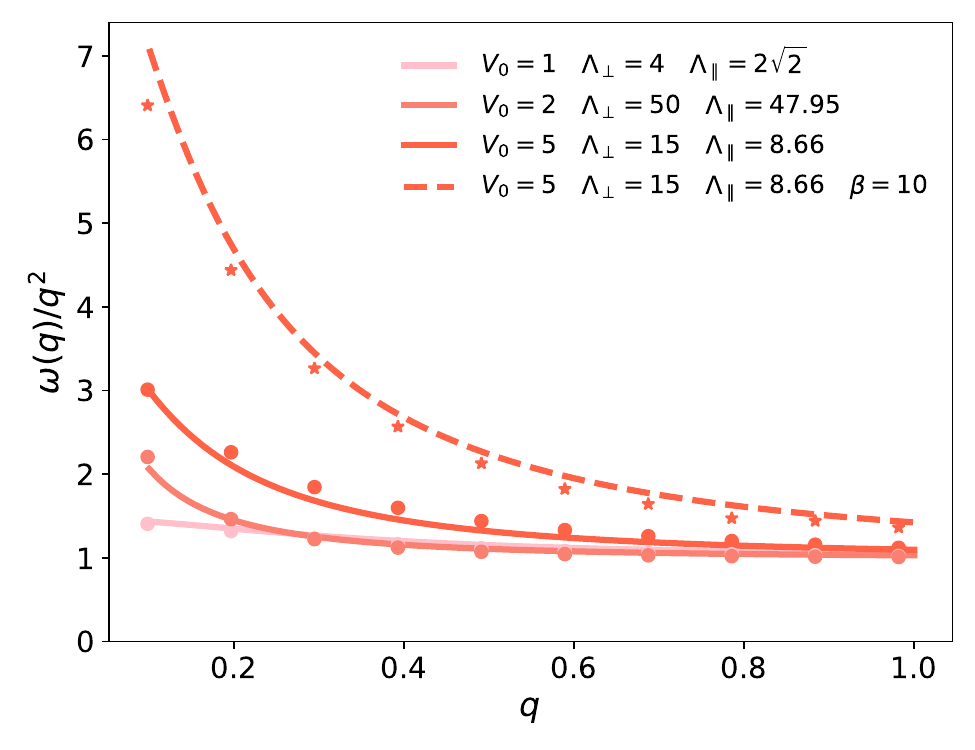}
    \label{fig:graphe2}
\end{subfigure}
\caption{Plot of $\omega(q)/q^2$ as a function of $q$ for various choices of the hydrodynamic lengthscales. Lines: Plot of the prediction in Eq.~\eqref{eq:dispersion-gore} for increasing values of $V_0$ and $\beta$, and various choices of the hydrodynamic lengthscales $\Lambda_{\perp,\parallel}$. Symbols: dispersion relation inferred from the relaxation of the interface in the numerical resolution of the  coupled partial differential equations Eqs.~\eqref{eq_phi} and \eqref{eq_v-enfonctiondephi}. Left: In the $q\xi\ll q\Lambda_{\perp,\parallel}\ll 1$ regime, the effective stiffness of the interface increases with $V_0$ and with $\beta$.  Right: In the $q\xi\ll 1\ll q\Lambda_{\perp,\parallel}$ regime, the same qualitative dependence on the parameters is observed, in addition to a crossover from one Edwards--Wilkinson regime to another.}
\label{fig:hydro-cross-EW}
\end{figure}

We now determine the dispersion relation in the same two limiting cases as discussed above. The overall phenomenology is identical. The effect of this additional term $B>0$ is  to systematically increase the relaxation rate, as shown in Fig.~\ref{fig:hydro-cross-EW}. In the $ q\xi,\,q\Lambda_{\parallel,\perp}\ll 1$ regime the dispersion relation becomes
\begin{equation}
    \begin{split}
        \frac{\omega(q)}{q^2}\underset{q\Lambda_{\perp},\,q\Lambda_{\parallel}\to 0}{\longrightarrow}&1-\frac{V_0}{2\int_u L\p_u\Phi}\int_u L(u) \partial_u\Phi(u) \int_{u'}\bigg\{\ee^{-|u-u'|/\Lambda_{\perp}}(u-u')\partial_{u'}\Phi(u')(1 + \beta \partial_{u'}^2 \Phi(u'))\\
        &-2\beta \ee^{-|u-u'|/\Lambda_{\perp}}(\partial_{u'} \phi(u'))^2\bigg\}
    \end{split}
\end{equation}
while in the other limit $ q\xi\ll 1\ll q\Lambda_{\perp,\parallel}$, we arrive at
\begin{equation}
    \begin{split}
        \frac{\omega(q)}{q^2}\underset{q\Lambda_{\perp},\,q\Lambda_{\parallel}\to \infty}{\longrightarrow}&1
    \end{split}
\end{equation}
When the wavelength becomes smaller than any hydrodynamic length scale (but still a bit larger than the width of the front) the dispersion relation becomes independent of all the hydrodynamic and mechanical details, as shown in Fig.~\ref{fig:hydro-cross-EW}(b).

\section{Without homeostatic pressure imbalance but with active forces}\label{sec:act}
\subsection{Nonlinear dynamics}
Following \cite{williamson_stability_2018} we consider the alternative mechanism where the nonequilibrium drive arises from a differential active force experienced by each species comprising the tissue, rather than from an imbalance in their homeostatic pressures. As we shall now see, this  changes the front selection mechanism which is neither pushed nor pulled. Instead, the front velocity is selected by the asymptotic behavior of the profile in the bulk of each tissue.\\

With equal division rates in both tissues the evolution equation for $\phi$ now loses its logistic term and reduces to an advection-diffusion equation:
\begin{equation}
    \p_t\phi+\bv\cdot\bnabla\phi=D\Delta \phi
    \label{eq_maitressse_active_force}
\end{equation}
Introducing active motility forces $f_i$ on the cells of tissue $i=A,B$ alters the dynamical balance equation. These active forces create an additional volume force that adds up to the existing stress
\begin{equation}
\gamma\bv=    \Delta f \phi \mathbf{e}_x  + \bnabla\cdot\bsigma 
    \label{eq_overdamped_active_forces}
\end{equation}
We have used the notation $\Delta f = f_A-f_B$. Equations~\eqref{eq_close_to_homeostasis} and \eqref{eq_incompressibility} can be combined to obtain the pressure field $P = -\frac{\bm \nabla \cdot \bv}{\kappa}$. Indeed, starting from Eq.~\eqref{eq_overdamped_active_forces} we arrive at
\begin{equation}
    \begin{split}
        T_{\alpha \beta} v_\beta= \frac{\Delta f}{\gamma} \phi \delta_{\alpha x} -  \frac{B}{\gamma} \left(
        \partial_{\beta}(\partial_{\beta} \partial_{\alpha}\phi)
        - \frac{1}{2}\partial_{\alpha}(\partial_{\gamma} \phi)(\partial_{\gamma} \phi)
        \right)
    \end{split}
\end{equation}
It is convenient to rescale the coordinates and fields as follows:
\begin{equation}
    t'=t/t_0,\,\, x'=x/\ell_{\perp},\,y'=y/\ell_{\perp},\,\,\, \bv'(x',y',t')=\bv(x,y,t) t_0/\ell_{\perp}
    \label{eq_dimensionless2}
\end{equation}
with $t_0=\frac{1 + \eta \kappa}{D\gamma \kappa}$, $\ell_{\perp}=\sqrt{\frac{1 + \eta \kappa}{\gamma \kappa}}$. We further introduce $\ell_{\parallel}=\sqrt{\frac{\eta}{\gamma}}$, $\tilde{\beta} = \frac{B\kappa}{D(1+\eta \kappa)}$ and $V_f = \frac{\Delta f t_0}{\gamma \ell_{\perp}} = \frac{\Delta f \sqrt{1 + \eta \kappa}}{\gamma^{3/2} D \sqrt{ \kappa}}$. Upon dropping the primes and using $\lambda = \frac{\ell_{\parallel}}{\ell_{\perp}}$ ($\lambda<1$) we obtain
\begin{equation}\label{eq_phi_active_force}
    \begin{split}
        \partial_t \phi  + \bv \cdot \bm \nabla \phi = \Delta \phi
    \end{split}
\end{equation}
The linear equation for $\bv$ Eq.~\eqref{V_eq} takes the form
\begin{equation}\label{eq_v-enfonctiondephi_active_force}
    \begin{split}
        \begin{pmatrix}
        1 - \partial_x^2 - \lambda^2\partial_y^2
        &
        -\left(1 -\lambda^2\right)\partial_x\partial_y
        \\
        -\left(1 -\lambda^2\right)\partial_x\partial_y
        &
        1 - \partial_y^2 - \lambda^2\partial_x^2
        \end{pmatrix}
        \bv =&  V_f \phi \mathbf{e}_x - \tilde{\beta}\bm \nabla \phi \Delta \phi
    \end{split}
\end{equation}
In what follows, for simplicity, we shall ignore the contribution from the Ericksen stress tensor (whose stabilizing effect is well understood). We begin by showing how the front selection mechanism differs from that discussed in \cite{ranft2014mechanically}.
Setting $\tilde{\beta} = 0$  and searching for a flat profile where the $y$-dependence of the field is dropped and where $\bv=v_x\be_x$ we arrive at the following coupled equations 
\begin{equation}
    \begin{split}
        &\partial_t \phi + v_x\partial_x \phi = \partial_x^2 \phi\\
        &v_x- \partial_x^2 v_x = V_f \phi
    \end{split}
\end{equation}
We are now dealing with a Burgers-like advection-diffusion equation for $\phi$, instead of a front selection emanating from nonconserved dynamics. Indeed,  searching for a front profile $\phi(x,y,t)=\Phi(x-ct)$, $v_x(x,y,t)=V(x-ct)$ leads to the coupled equations
\begin{equation}\label{eq:mfprofiles}
    -(c-V)\Phi'=\Phi'',\,\,V-V''=V_f\Phi
\end{equation}
which we integrate over the whole real axis (using the boundary conditions $\Phi'(\pm\infty)=0$):
\begin{equation}
    c = - \int_{-\infty}^{+\infty} \dd u V(u)\Phi'(u)
\end{equation}
With the use of $V_f\Phi'=V'-V'''$ and an integration by parts we see that
\begin{equation}
    c = - V_f^{-1}\int_{-\infty}^{+\infty} \dd u V(u)(V'(u)-V'''(u))=-\frac{1}{2V_f}\left[V^2\right]_{-\infty}^{+\infty}-\frac{1}{V_f}\left[VV''\right]_{-\infty}^{+\infty}+\frac{1}{2V_f} \left[V'^2\right]_{-\infty}^{+\infty}=\frac{V_f}{2}
\end{equation}
The coupling to the auxiliary hydrodynamic field does not alter the selected velocity with respect to that predicted by a Burgers equation with a nonlinearity amplitude $V_f$. This is the same velocity as that predicted with the linearized approach adopted in \cite{williamson_stability_2018}, which was also observed in Fig.~3(a) of \cite{buscher_instability_2020} (with a perfect linear fit). Our numerical resolution of Eqs.~\eqref{eq_phi_active_force} and \eqref{eq_v-enfonctiondephi_active_force} with $\beta=0$ is also consistent with $c=V_f/2$. The front is now neither pushed nor pulled. This is a traveling Burgers wave. We are now in a position to investigate the stability of the flat interface along the $y$ transverse direction.

\subsection{Dispersion relation for weak deformations}

\begin{figure}[H]
    \centering
\includegraphics[width=0.495\linewidth]{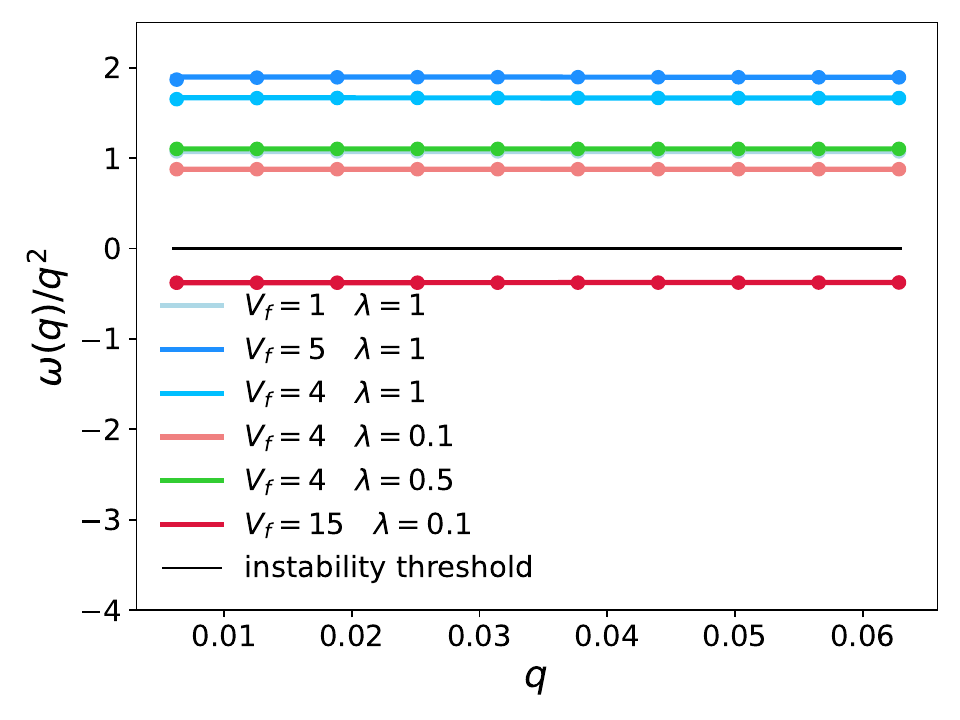}\begin{overpic}[width=0.495\linewidth]{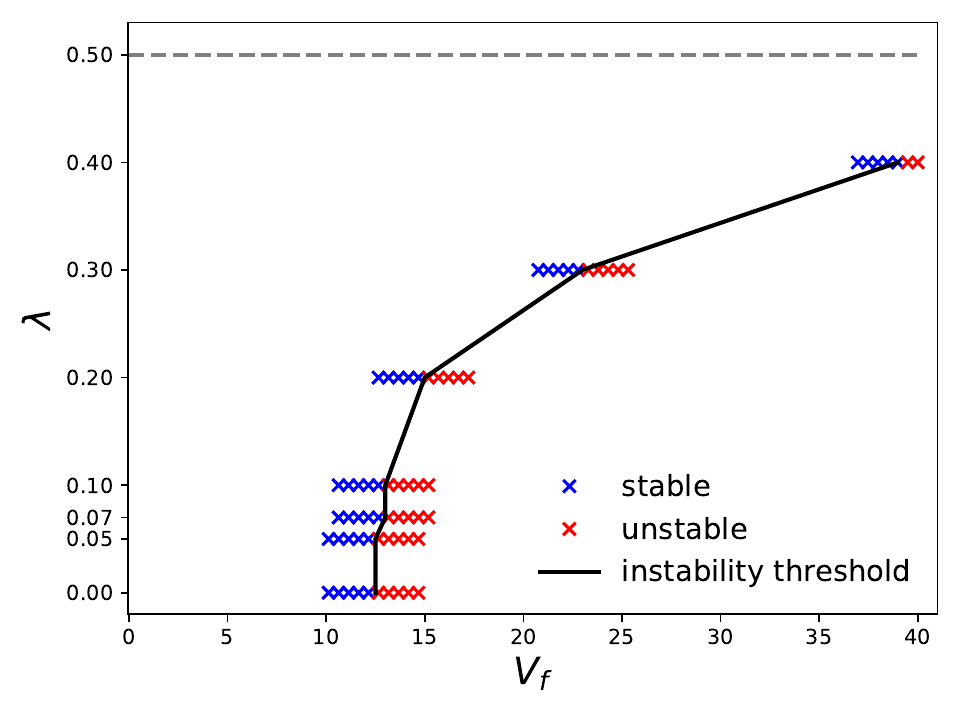}
\put (70,40) {\textcolor{red}{unstable}}
\put (25,40) {\textcolor{blue}{stable}}
\end{overpic}
    \caption{Left: Plot of $\omega(q)/q^2$ given by \eqref{eq_dispersion_active}, for  various values of $V_f$ and $\lambda$, determined using the numerical solution of Eq.~\eqref{eq:mfprofiles} (continuous lines).    The bullets are obtained by a direct measurement of the interface in a numerical resolution of Eqs.~\eqref{eq_phi_active_force} and \eqref{eq_v-enfonctiondephi_active_force} and excellent agreement is achieved. Right: The region of stability of the interface in the $(V_f,\lambda)$ plane was determined by scanning the parameter space  (crosses) in the vicinity of the black curve. The dashed line at $\lambda=0.5$ is a guide for the eye and stands for the upper value above which no instability is predicted at asymptotically large $V_f$.}
    \label{fig:placeholder}
\end{figure}

For weak deformations of the interface, we can linearize Eqs.~\eqref{eq_phi_active_force} and \eqref{eq_v-enfonctiondephi_active_force} after plugging the form Eq.~\eqref{eq_deviations}. In the same way as in subsection~\ref{subsec:homeo}, this leads to three linear yet coupled equations involving the four unknowns $h$, $\delta\phi$ and $\delta\bv$, which we write in Fourier space for the $y$ variable:
\begin{equation}
    \begin{split}
    \left\{\begin{array}{ll}
        &-\partial_t h \partial_u \Phi + \partial_t \delta\phi =  (c-V)\partial_u \delta\phi - \delta v_u \partial_u\Phi + (-q^2+\p_u^2) \delta \phi +  q^2 h \partial_u\Phi \\
        \\
         &\bigg(1 -\partial_u^2 + \lambda^2 q^2 \bigg) \delta v_u  - \left(1 - \lambda^2\right)iq\partial_u \delta v_y  =  V_f \delta \phi + \lambda^2 q^2 h \partial_u V\\
        \\
         &\bigg(1 -\lambda^2 \partial_u^2 +q^2 \bigg) \delta v_u  - \left(1 - \lambda^2\right)iq\partial_u \delta v_u  =-(1-\lambda^2)iq h \partial_u^2 V  \end{array}  \right. 
    \end{split}
\end{equation}
We are now in a position to implement the exact same method as that described after Eq.~\eqref{eq_operator}. We do not repeat the mathematical steps (see appendix~\ref{subsec:app-gamma}) and go directly to the dispersion relation:
\begin{equation}
    \begin{split}
        \omega(q)=&q^2-\frac{V_f}{2}\int_u\p_u \Phi(u) \int_{u'}  \bigg\{\bigg( \ee^{-|u-u'|} + q^2\ee^{-|u-u'|\sqrt{1+(\lambda q)^2}/\lambda}\\
        &- (1+q^2)\ee^{-|u-u'|\sqrt{1+q^2}}\bigg)\text{sign}(u-u')\Phi(u')\bigg\}
    \end{split}
\label{eq_dispersion_active}
\end{equation}
In the limit of large wavelength we recover an Edwards-Wilkinson scaling, namely
\begin{equation}
    \begin{split}
    \frac{\omega(q)}{q^2} \simeq 1 - \frac{V_f}{2}\int_{u,u'} \partial_u\Phi(u) \Phi(u')\left[\left(\ee^{-|u-u'|/\lambda}-\ee^{-|u-u'|}\right)\text{sign}(u-u')+\frac 12 (u-u')\ee^{-|u-u'|}\right]
    \end{split}
\end{equation}
Fortunately, the profiles $\Phi$ and $V$ (which are independent of $\lambda$) can be understood in simple mathematical terms, since not only the $V_f\to 0^+$ and $V_f\to+\infty$ limiting cases, but also the particular $V_f=4$ case, can be worked out analytically ($V_f\to 0^+$ means a very soft front while $V_f\to+\infty$ means a very steep front). For $V_f=4$ we find that $\omega(q)/q^2$ monotonously increases as a function of $\lambda\in]0,1[$ from $\frac{13}{15}$ up to $\frac 53$. The $V_f\to+\infty$ limit, for which the profile reduces to a step function, can be worked out exactly:
\begin{equation}
    \frac{\omega(q)}{q^2}\simeq\frac{2\lambda-1}{4}V_f
\end{equation}
This tells us that a large-wavelength instability indeed sets in for $V_f$ large enough and $\lambda$ small enough. In Fig.~\ref{fig:placeholder} we see that the instability appears in the large $V_f$ and small $\lambda$ regime. The ratio $\lambda=\sqrt{\frac{\eta\kappa}{1+\eta\kappa}}$ is small when $\eta\ll\kappa^{-1}$, namely when the hydrodynamic viscosity is small compared to the effective bulk viscosity for cell number change (the tissue loses its elastic character in that regime). It is then unable to regulate density fluctuations which follow the velocity in a way reminiscent of the Saffman-Taylor instability discussed in \cite{williamson_stability_2018}. 

Direct comparison  with either the numerical simulations of \cite{buscher_instability_2020} or with the analysis of \cite{williamson_stability_2018} is delicate. In the latter nonlinear phenomena have been simplified and we have shown that introducing an {\it ad hoc} surface tension, while physical, does not allow to fully understand the domain of stability of the interface shown in Fig.~\ref{fig:placeholder}(b). In the former, we can confirm that in a weak drive regime the interface is stable and that it takes a large enough $V_f$ to generate a $\lambda$-dependent instability. Finally, note that we have worked in the $B=0$ limit: restoring the Ericksen stress tensor contribution is likely to shrink the instability region.

\section{Concluding remarks}

Starting from the continuous hydrodynamic description of \cite{ranft2014mechanically}, we have been able to build from first principles a dispersion relation for the relaxation of the interface between two tissues. The interface is described as an emerging collective coordinate living at the same mesoscopic scale as the hydrodynamic fields themselves. Based on the methods of \cite{sarfati_bulk_2026}, our results show that, at least in the absence of a compressibility mismatch, an interface driven by distinct homeostatic pressures is always stable. This is fully consistent with the numerical simulations in \cite{buscher_instability_2020}. 

The relaxation spectrum always exhibits the Edwards-Wilkinson scaling except in a crossover region where the wavelength becomes of the same order as the hydrodynamic length scale. Indeed, we have shown the existence of two distinct Edwards-Wilkinson regimes, depending on how hydrodynamic length scales compare with the wavelength of the perturbation.

We have extended the hydrodynamic description to a weak compressibility mismatch and have found no corresponding instability as we present in~\ref{subsec:app-epsilon} (perhaps such an instability as discussed in \cite{williamson_stability_2018, buscher_instability_2020} cannot be captured by a hydrodynamic description). In addition, we have explored the fate of an interface driven by tissue-dependent active forces and we have found the physics of the front selection mechanism to differ from that at work in the case of a homeostatic pressure imbalance. The front velocity is directly proportional to the imbalance in active forces. We have found the interface to be unstable against small perturbations for large enough drive, in a viscosity dependent fashion.\\
In our analysis, following~\cite{williamson_stability_2018,buscher_instability_2020}, we have cautiously disentangled homeostatic pressure imbalance from active force imbalance. When combined, the nonlinear front selection mechanism, and the accompanying interface stability analysis, would involve exploring the competition between the nonconservative pushed-pulled {\it vs.} advection-diffusion mechanisms. This is both an arduous piece of mathematical literature~\cite{sanchez1997travelling,hendersonslowfast2022} and an experimentally relevant setting beyond tissues~\cite{cochet2021hypoxia} that is certainly worth future explorations.

\ack {We} thank J. Ranft for a useful exchange on the two-dimensional hydrodynamic formulation and G. Salbreux for an interesting discussion in the early stage of this work.

\section*{Appendix}
\subsection{Explicit expressions of the operators entering the derivation of the dispersion relation  }\label{subsec:app-gamma}

Starting from 

\begin{equation}
    \Gamma(\delta \phi -h \partial_u\Phi) = \Gamma' h \partial_u \Phi
\end{equation}

with 

\begin{equation}
    \Gamma = \partial_t + \Omega + F(q,u)
\end{equation}

where the lowest eigenvalue of $\Gamma$ is called $\Lambda_0$ which reads in the low-q regime 

\begin{equation}
    \Lambda_0(q) = \frac{\langle L,F(q,u)\partial_u\Phi \rangle}{\langle L,\partial_u\Phi \rangle}
\end{equation}

\begin{equation}
    \Gamma' = \Tilde{F}(q,u) - F(q,u)
\end{equation}

The dispersion relation reads

\begin{equation}
    \begin{split}
        \omega(q) =& \Lambda_0(q) + \frac{\langle L,\Gamma'\partial_u\Phi \rangle}{\langle L,\partial_u\Phi \rangle}\\
        =&\frac{\langle L,\tilde{F}(q)\partial_u\Phi \rangle}{\langle L,\partial_u\Phi \rangle}
    \end{split}
\end{equation}

Let us recall 
\begin{equation}\label{eq_Green}
    \mathcal{G}(q) =
        \begin{pmatrix}
         \frac{\ee^{-\frac{|u|}{\zeta_{\perp}(q)}}}{\zeta_{\perp}(q)} - q^2 \zeta_{\parallel}(q) \ee^{\frac{-|u|}{\zeta_{\parallel}(q)}}
        &
        -iq \text{ sign}(u)\bigg(\ee^{-\frac{|u|}{\zeta_{\perp}(q)}} - \ee^{\frac{-|u|}{\zeta_{\parallel}(q)}}\bigg)
        \\[6pt]
        -iq \text{ sign}(u)\bigg(\ee^{\frac{-|u|}{\zeta_{\perp}(q)}} - \ee^{\frac{-|u|}{\zeta_{\parallel}(q)}} \bigg)
        &
         \frac{\ee^{\frac{-|u|}{\zeta_{\parallel}(q)}}}{\zeta_{\parallel}(q)} - q^2 \zeta_{\perp}(q) \ee^{\frac{-|u|}{\zeta_{\perp}(q)}}
        \end{pmatrix} 
\end{equation}

\subsubsection{In the absence of the Ericksen stress tensor}

In what follows ($.$) indicates that the convolution applies to the function the operator is applied. Let us define $g_{11}(u,q) = \mathcal{G}_{11}(u,q) - \mathcal{G}_{11}(u,0)$

\begin{equation}
    \begin{split}
        \left\{\begin{array}{ll}
        F(q,u) &= q^2 - 2\Lambda_{\perp}V_0 \partial_u\Phi(u)\bigg(\big(g_{11}(u,q)*\partial_u .\big) + iq\big(\mathcal{G}_{12}(u,q)*.\big)\bigg)\\
        \\
        \tilde{F}(q,u) =& q^2 - 2\Lambda_{\perp}V_0 \partial_u\Phi(u)\bigg(\big(\partial_u g_{11}(u,q)*\partial_u \Phi(u)\big) + iq\big(\mathcal{G}_{12}(u,q)*\partial_u\Phi(u)\big)\bigg)\\
        \\
        \Omega =& -\partial_u^2 + (V(u) - c)\partial_u -(1-2\Phi(u)) - 2\Lambda_{\perp}\partial_u\Phi(u)( \mathcal{G}_{11}(u,0) * \partial_u.)
        \end{array}  \right.
    \end{split}
\end{equation}

\subsubsection{Including cell differential adhesion}

Let us define $E = \partial_u^2\Phi(u)\partial_u + \partial_u\Phi(u)\partial_u^2$

\begin{equation}
    \begin{split}
        \left\{\begin{array}{ll}
        F(q,u) =& q^2 - 2\Lambda_{\perp}V_0 \partial_u\Phi(u)\bigg(\big(g_{11}(u,q)*(\partial_u + \beta E ).\big) + iq\big(\mathcal{G}_{12}(u,q)*(1+\beta \partial_u^2 \Phi(u)).\big)\\
        &- \beta q^2 \big(\mathcal{G}_{11}(u,q)*\partial_u\Phi(u) .\big)\bigg)
        \\
        \\
        \tilde{F}(q,u) =& q^2 - 2\Lambda_{\perp}V_0 \partial_u\Phi(u)\bigg(\big(\partial_u g_{11}(u,q)*\partial_u \Phi(u)(1+\beta\partial_u^2\Phi(u))\big) \\
        &+ iq\big(\mathcal{G}_{12}(u,q)*\partial_u\Phi(u)(1+\beta\partial_u^2\Phi(u))\big) -\beta q^2\big(\mathcal{G}_{11}(u,q)*(\partial_u\Phi(u))^2\big)\bigg)
        \\
        \\
        \Omega =& -\partial_u^2 + (V(u) - c)\partial_u -(1-2\Phi(u)) - 2\Lambda_{\perp}\partial_u\Phi(u)\big( \mathcal{G}_{11}(u,0) * (\partial_u + \beta E).\big)
        \end{array}  \right.
    \end{split}
\end{equation}

\subsubsection{Without homeostatic pressure imbalance but with active forces}

\begin{equation}
    \begin{split}
    \left\{\begin{array}{ll}
        F(q,u) =& q^2 + V_f \partial_u\Phi(u) \big(g_{11}(u,q)*.\big)
        \\
        \\
        \tilde{F}(q,u) =& q^2 + V_f \partial_u\Phi(u) \big(\partial_u g_{11}(u,q)*\Phi(u)\big)\\
        \\
        \Omega =& -\partial_u^2 + (V(u)-c)\partial_u + V_f\partial_u\Phi(u)\big(\mathcal{G}_{11}(u,0)*.\big)
         \end{array}  \right.
    \end{split}
\end{equation}

\subsection{Effect of a differential compressibility}
\label{subsec:app-epsilon}
\begin{figure}
\begin{subfigure}{0.495\textwidth}
    \centering
    \includegraphics[width=\linewidth]{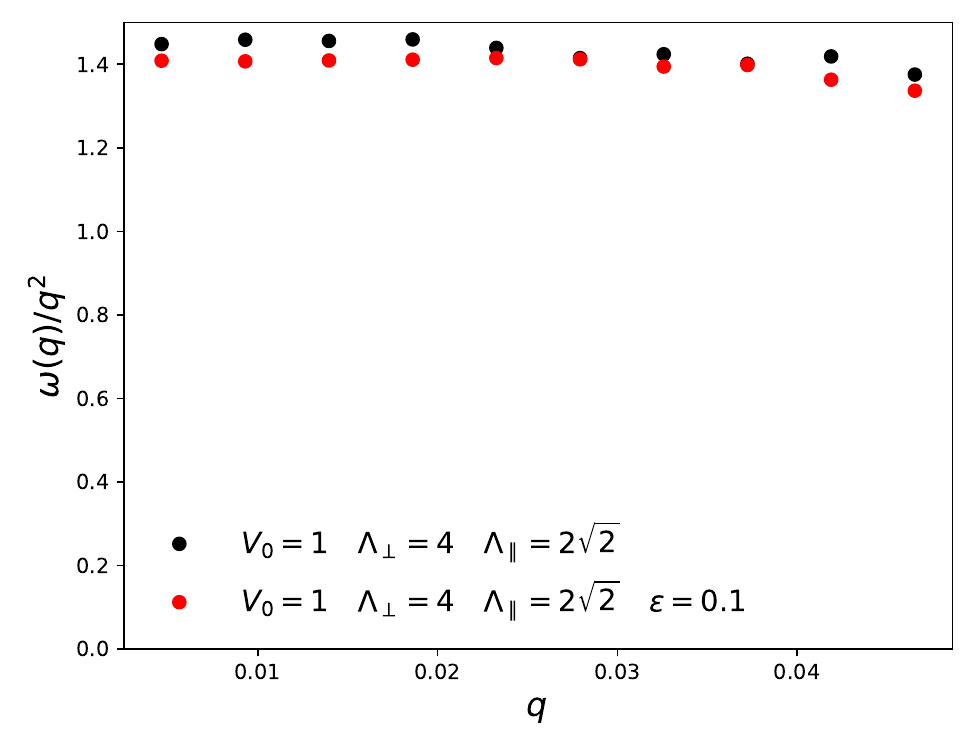}
    \label{fig:graphe1_eps}
\end{subfigure}
\hfill
\begin{subfigure}{0.495\textwidth}
    \centering
    \includegraphics[width=\linewidth]{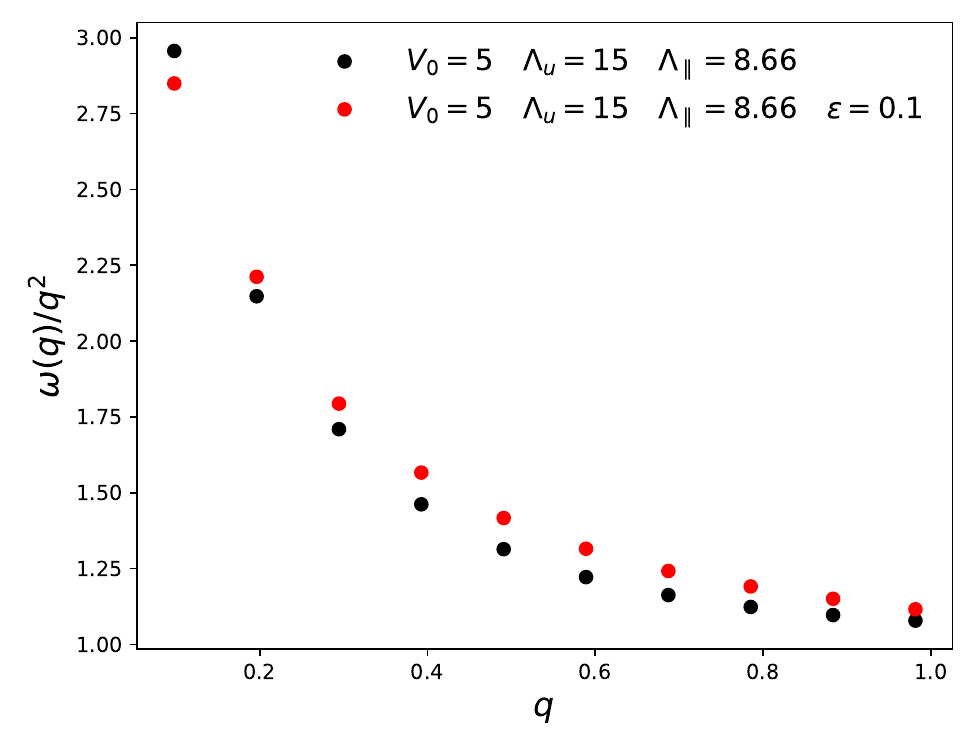}
    \label{fig:graphe2_eps}
\end{subfigure}
\caption{Plot of $\omega(q)/q^2$ as a function of $q$. Left and right plot are done respectively in the $q\Lambda_{\perp,\parallel} \ll 1$ and $q\Lambda_{\perp,\parallel}q \gg 1 $ regimes. Dots are numerical solutions of equations \eqref{eq_phieps},\eqref{eq_v-enfonctiondephieps}. The solution at equal compressibilities is in black while the solution at $\eps>0$ is in red.}
\label{fig_eps}
\end{figure}

We now allow for the compressibilities in tissues $A$ and $B$ to be different, namely $\kappa_A \neq \kappa_B$.
\begin{equation}
    k_i \simeq \kappa_i \left( P_i^{h} - P \right)
    \label{eq_close_to_homeostasis_eps}
\end{equation}
We prepare for our theoretical analysis by denoting 
$\kappa_A = \kappa$ and $\kappa_B = \kappa -\Delta \kappa $, and by introducing $\eps = \frac{\Delta \kappa}{\kappa}$. The deviations of the pressure with respect to its homeostatic value are assumed to be small, which, for consistency, also requires the coefficients $\kappa_i$ to remain close to each other, or, equivalently, that we work in the small $\eps$ regime:
\begin{equation}
    \begin{split}
        P_B^h - P =& \frac{\frac{\bm \nabla \cdot \bm v}{\kappa} - \phi \Delta P^h}{1+\eps(\phi-1)}
    \end{split}
\end{equation}
To  first order in $\eps$, the overdamped equation of motion takes the form 
\begin{equation}
    \begin{split}
        T_{\alpha\beta}v_{\beta} + \frac{\eps}{\kappa \gamma}\bigg(\left(\phi-1\right)\partial_{\alpha}\partial_{\beta}v_{\beta}  + \partial_{\alpha}\phi \partial_{\beta}v_{\beta} \bigg) =& -\frac{\Delta P^h}{\gamma} \partial_{\alpha}\phi +  \frac{\eps\Delta P^h}{\gamma} \partial_{\alpha}\phi\left(1-2\phi\right) \\
        &-  \frac{B}{\gamma} \left(
        \partial_{\beta}(\partial_{\beta} \partial_{\alpha}\phi)
        - \frac{1}{2}\partial_{\alpha}(\partial_{\gamma} \phi)(\partial_{\gamma} \phi)
        \right)
    \end{split}
    \label{V_eq}
\end{equation}
where the hydrodynamic interaction tensor $T_{\alpha\beta}$ is defined in Eq.~\eqref{eq_tenseur_T}. Now that the compressibilities are distinct in Eq.~\eqref{eq_close_to_homeostasis} one can rewrite Eq.~\eqref{eq_maitressse} to order one in $\eps$:
\begin{equation}\label{eq_phieps}
    \begin{split}
        \partial_t \phi  + {\bf v} \cdot \bm \nabla \phi = \Delta \phi  + \phi\left(1-\phi\right)\big(1 + \eps\left(\bm \nabla \cdot {\bf v} - \phi \right)\big)
    \end{split}
\end{equation}
The linear equation for $\bv$ in Eq.~\eqref{V_eq} takes the form
\begin{equation}\label{eq_v-enfonctiondephieps}
    \begin{split}
        \begin{pmatrix}
        1 - \Lambda_\perp^2\partial_x^2 - \Lambda_\parallel^2\partial_y^2
        &
        -\left(\Lambda_\perp^2 -\Lambda_\parallel^2\right)\partial_x\partial_y
        \\
        -\left(\Lambda_\perp^2 -\Lambda_\parallel^2\right)\partial_x\partial_y
        &
        1 - \Lambda_\perp^2\partial_y^2 - \Lambda_\parallel^2\partial_x^2
        \end{pmatrix}
        {\bf v} =&  -2\Lambda_\perp V_0 \bigg[\bm \nabla \phi \bigg(1+ \beta \Delta \phi \\+
        &\eps \left(1 - 2\phi + \bm \nabla \cdot {\bf v} \right)\bigg)  - \eps\left(1 - \phi\right)\bm \nabla(\bm\nabla \cdot \bm v)\bigg) \bigg]
    \end{split}
\end{equation}
Because the operator acting on $\bv$ in  Eq.~\eqref{eq_v-enfonctiondephieps} is not translationally invariant anymore, one can again split the fields as in Eq.~\eqref{eq_deviations},  but inverting the operator acting on $\delta \bv$ is not analytically tractable. One can however still solve numerically the coupled system of partial differential equations in  \eqref{eq_phieps} and \eqref{eq_v-enfonctiondephieps}. Since only small values of $\eps$ are allowed in this description, one cannot push the differences in compressibility enough so that we can explore a strong effect on the interface dynamics. The only controlled statement we can make is that  the presence of a small difference in compressibilities tends to moderately reduce the effective surface tension as shown in Fig.~\ref{fig_eps}, where the red curve at $\eps>0$ indeed sits a little below that at $\eps=0$.

\subsection{Numerical solutions of the partial differential equations}

The system of coupled partial differential equations of Eqs.~\eqref{eq_phi} and \eqref{eq_v-enfonctiondephi} is solved using a Fourier pseudospectral method implemented in the Dedalus framework~\cite{Dedalus_PhysRevResearch.2.023068}, with periodic boundary conditions in both spatial directions. Time integration is performed using a second-order semi-implicit backward differentiation formula (also known as SBDF2). We first solved the equations in one space dimension to obtained the steady profile of the front. We also numerically solve for the eigenvector with zero eigenvalue of the operator describing the linearized dynamics. We use these in Eq.~\eqref{eq:dispersion-gore} to come up with a numerical estimate of the dispersion relation. Then we test this theoretical description by looking at the relaxation time of the relaxed interface. The initial conditions are constructed from a one-dimensional, relaxed front profile, which is extended in the transverse direction by introducing a sinusoidal modulation of the front position. In practice, the initial scalar density field is defined as $\phi(x,y,0) = \Phi(x-A \cos(qy))$  with $A$ being the small amplitude of the perturbation and $q=\frac{2p\pi}{L_y}$ the wave vector for which the dynamics is investigated.
For the $ q\Lambda_{\perp},q\Lambda_{\parallel}\ll 1$ regime the length of the simulation box along the flat interface direction is taken to be $L_y = 1350$ whereas for the $ \,q\Lambda_{\perp},\,q\Lambda_{\parallel}\gg 1$ regime $L_y=64$.  The simulation box in the moving front direction is taken to be    $L_x \propto c t_{\text{f}}$  where $t_{\text{f}}$ is the total simulation time, which is chosen such that we can observe the exponential relaxation of the interface. The discretization step in the $y$-direction is set to be $dy = 1$ for $L_y = 1350$ and $dy=0.25$ for $L_y=64$. The time step $dt$ is adapted to the value of $V_0$ and $\beta$ to preserve numerical stability and is taken from $dt=0.01$ for $V_0=5$, $\beta=10$ up to $dt=0.1$ for $V_0 = 1$, $\beta =0$. The spatial discretization step along the $x$-direction is always $dx = 0.25$, the front of width $1$ is then resolved with 4 grid points.\\

The location $h(y,t)$ of the interface  is identified as the $\phi = 0.5$ contour line. For each $y$-position and time snapshot, we locate the changes of sign of $\phi(x,y,t)-0.5$ along the $x$-direction. Around each detected sign change, the local interface position is determined by fitting a cubic polynomial to four neighboring grid points and computing its real root within the corresponding interval. This procedure provides us with the interface deformation $h(y,t)$ over the entire transverse direction. The agreement between this pragmatic numerical definition of $h$ and the formal one in Eq.~\eqref{eq_definterface} is remarkable (as can be seen in Figs.~\ref{fig:hydro-cross-EW} and \ref{fig:placeholder}).

\bibliographystyle{plain}
\bibliography{biblio}

\end{document}